\documentclass[a4paper,11pt]{article}
\usepackage{pos}

\long\def \blockcomment #1\endcomment{}

\def\ttl#1{{\it #1}}

\title{Dilaton chiral perturbation theory and applications}
\ShortTitle{Dilaton chiral perturbation theory}

\author*[a]{Maarten Golterman}
\author[b]{Yigal Shamir}

\affiliation[a]{Department of Physics and Astronomy, San Francisco State University,\\
San Francisco, CA 94132, USA}

\affiliation[b]{Raymond and Beverly Sackler School of Physics and Astronomy,\\
Tel~Aviv University, 69978, Tel~Aviv, Israel}

\emailAdd{maarten@sfsu.edu}
\emailAdd{shamir@tauex.tau.ac.il}

\abstract{We review dilaton chiral perturbation theory (dChPT),
the effective low-energy theory for the light sector of near-conformal,
confining theories. dChPT
provides a systematic expansion in both the fermion mass
and the distance to the conformal window.
It accounts for the pions and the light scalar, the approximate
Nambu--Goldstone bosons for chiral and scale symmetry, respectively.
A unique feature of dChPT is the existence of
a large-mass regime
in which the theory exhibits approximate hyperscaling,
while the expansion nevertheless remains systematic.
We discuss applications to lattice data, presenting successes
as well as directions for future work.
}

\FullConference{%
 The 38th International Symposium on Lattice Field Theory, LATTICE2021
  26th-30th July, 2021
  Zoom/Gather@Massachusetts Institute of Technology
}

\begin{document}
\maketitle

\section{Introduction}

\begin{figure}[t]
\begin{center}
\includegraphics*[height=6cm]{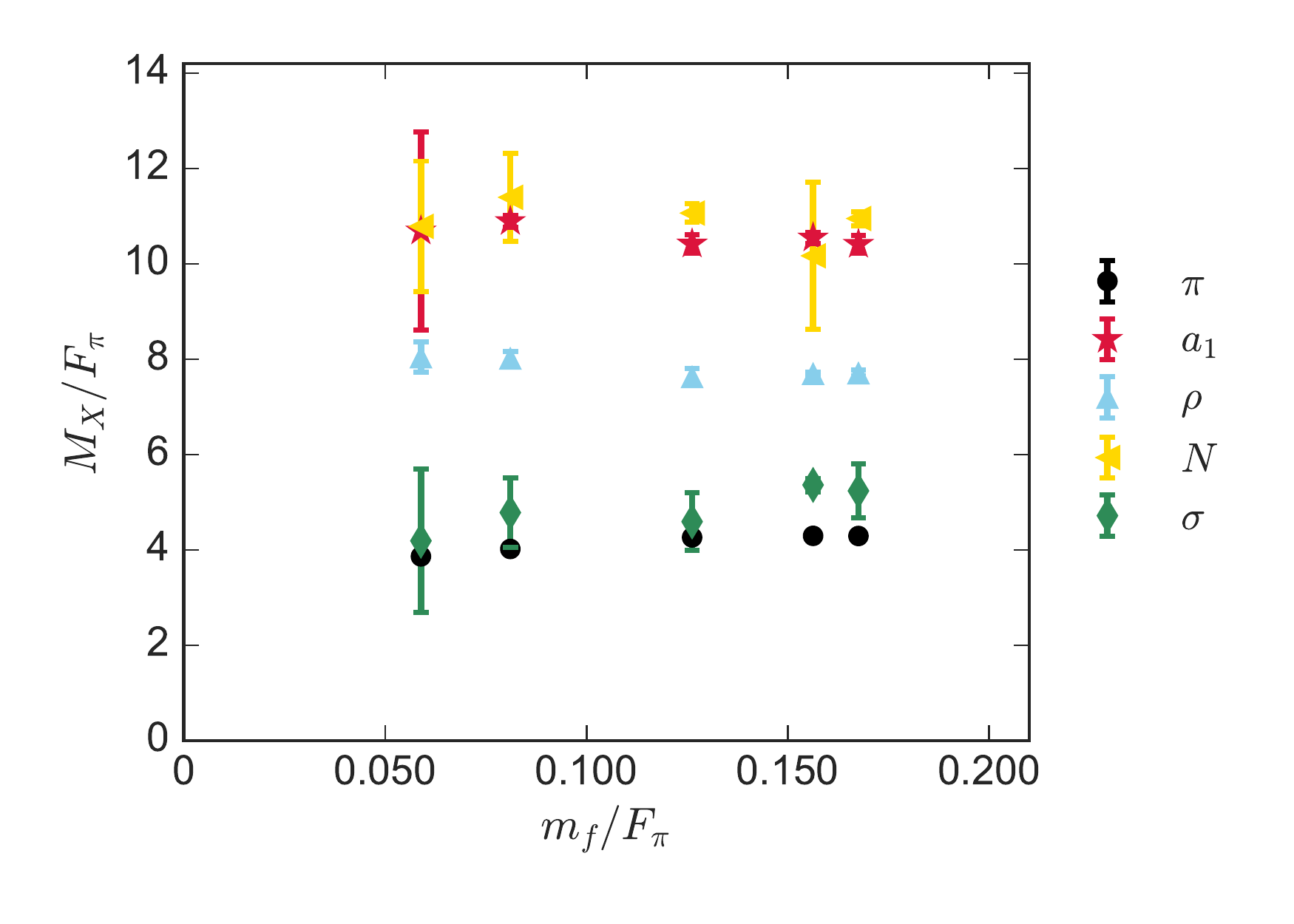}
\end{center}
\vskip-3ex
\caption{\it Hadron masses as a function of fermion mass,
all in units of $F_\pi$, in the SU(3) gauge theory with 8 fundamental flavors,
from Ref.~\cite{LSD}.
}
\label{fig1}
\end{figure}

Let us start with a look at spectral data for the SU(3) gauge theory
with $N_f=8$ fundamental fermions obtained by the LSD collaboration
\cite{LSD0,LSD}, shown in Fig.~\ref{fig1}.
In comparison with QCD, several salient differences stand out:
First, we note the presence of a stable $0^{++}$ state,
denoted by the points labeled $\sigma$, with a mass virtually
degenerate with the pions, over this mass range.   Second, within errors,
the spectrum shows hyperscaling:  hadronic mass ratios are
essentially independent of the fermion mass.
In addition, pion ``taste splittings''
associated with the use of staggered fermions in all these simulations,
{\it i.e.}, non-degeneracies in the pion spectrum due to scaling violations,
behave very differently as a function of the fermion mass
in comparison with QCD (not shown in the figure).
Similar results have been
found in Ref.~\cite{KMI} by the LatKMI collaboration for the same theory, and in the SU(3) theory with two sextet fermions by the
LatHC collaboration \cite{LatHC,LatHC2}, or with four light and six \cite{LSD10} or eight \cite{BHRWW} heavy fundamental fermions.   In this talk, we will discuss to what extent these data are decribed by tree-level dilaton
Chiral Perturbation Theory (dChPT).

After a brief review of dChPT and its lowest-order lagrangian, we will
explain the existence of
a ``large-mass'' regime in which dChPT predicts approximate hyperscaling,
as seen in Fig.~\ref{fig1}.  Going beyond hyperscaling, we will then
apply dChPT to fit the LSD and LatKMI data,
including the measured staggered taste splittings.
We conclude with a brief discussion of what might be next.

\section{Lowest-order dChPT}

Dilaton ChPT is based on the following set of assumptions \cite{GS,GSlat16}:
\begin{itemize}
\item
Every gauge theory below the conformal sill
contains Nambu--Goldstone bosons, or ``pions,''
associated with the spontaneous breaking of chiral symmetry.
The pions become massless in the limit $m\to 0$,
where $m$ is the (degenerate) fermion mass.
\item Scale invariance gets restored in the infrared as we approach the conformal
window, with the trace anomaly being proportional to the distance to the conformal window.
In the Veneziano limit $N_f\to\infty$, $N_c\to\infty$, with $n_f = N_f/N_c$ fixed, this happens
when $n_f$ approaches a critical value $n_f^*$
from below.  The difference $n_f-n_f^*$
is a new small parameter.
\item The theory contains a {\it dilaton}, {\it i.e.}, a Nambu--Goldstone boson associated with scale symmetry breaking, which becomes massless in the
double  limit $n_f-n_f^*\to 0$ and $m\to 0$.
\item In addition, some technical assumptions on the dilaton potential are needed, see Ref.~\cite{GSlarge}.
\end{itemize}
With these assumptions, one can prove that a systematic power counting in the
small parameters
\begin{equation}
\label{powerc}
p^2\sim m\sim n_f-n_f^*\sim 1/N_c
\end{equation}
exists, and construct the lowest-order (${\cal{O}}(p^2)$) lagrangian,
\begin{eqnarray}
\label{lag}
\cal{L}&=&\frac{1}{4}f_\pi^2e^{2\tau}\,\mbox{tr}(\partial_\mu\Sigma^\dagger\partial_\mu\Sigma)+\frac{1}{2}f_\tau^2 e^{2\tau}\partial_\mu\tau\partial_\mu\tau\\
&&-\frac{1}{2}f_\pi^2 B_\pi m e^{(3-\gamma_*)\tau}\mbox{tr}(\Sigma+\Sigma^\dagger)+f_\tau^2 B_\tau e^{4\tau}c_1\left(\tau-\frac{1}{4}\right)\ .\nonumber
\end{eqnarray}
Here $\Sigma=\mbox{exp}(2i\pi/f_\pi)$ is the non-linear pion field and
$\tau$ is the dilaton field. $c_1$ is a parameter proportional to
$n_f-n_f^*$, of order  $p^2$ in our power counting.  At leading order, five
low-energy constants (LECs) appear: $f_{\pi,\tau}$, $B_{\pi,\tau}$,
and $\gamma_*$, the mass anomalous dimension at the infra-red fixed point
at the conformal sill.
For a detailed discussion of the lagrangian and the power counting,
see Ref.~\cite{GS}.  We have used
the freedom to shift the field $\tau$ such that $v(m)=\langle\tau\rangle$ vanishes for $m=0$, at this order.

The saddle point equation for the classical solution $v(m)$ is
\begin{equation}
\label{vm}
\frac{m}{c_1\cal{M}}=v(m)e^{(1+\gamma_*)v(m)}\ ,\qquad{\cal{M}}\equiv\frac{4f_\tau^2 B_\tau}{f_\pi^2 B_\pi N_f(3-\gamma_*)}\ .
\end{equation}
The tree-level pion and dilaton masses, and their decay constants, are,
using Eq.~(\ref{vm}),
\begin{eqnarray}
\label{treelevel}
M_\pi^2&=&2B_\pi me^{(1-\gamma_*)v(m)}=2B_\pi c_1{\cal{M}} v(m)e^{2v(m)}\ ,\\
M_\tau^2&=&4B_\tau c_1 e^{2v(m)}\left(1+(1+\gamma_*)v(m)\right)\ ,
\nonumber\\
F_{\pi,\tau}&=&f_{\pi,\tau} e^{v(m)}\ .\nonumber
\end{eqnarray}
Unlike in ordinary ChPT, here the tree-level decay constant(s) depend
on the fermion mass via the factor $e^{v(m)}$.

\section{The small- and large-mass regimes}
An important observation is that in our power counting
the ratio $\frac{m}{c_1\cal{M}}$ is parametrically ${\cal O}(1)$.
But this ratio can still be small,
corresponding to the {\it small-mass regime}, or large, corresponding to the
{\it large-mass regime}.

In the small-mass regime, $\frac{m}{c_1\cal{M}}\ll 1$, and $v\propto m$
while $e^v = 1+{\cal O}(m)$.
One has $M_\pi^2=2B_\pi m\ll M_\tau^2=4B_\tau c_1\propto |n_f-n_f^*|$.
Even though, relative to the pions, the dilaton decouples,
its mass still remains parameterically smaller than that of
all other hadrons. For energies below the scale set by the dilaton mass
the physics of pions only is described by standard ChPT.

The large-mass regime, $\frac{m}{c_1\cal{M}}\gg 1$, is more
interesting \cite{GSlarge}.   In Eq.~(\ref{vm}) the exponential
dominates, and this equation has the approximate solution
\begin{equation}
\label{evm}
e^{v(m)}\sim\left(\frac{m}{c_1\cal{M}}\right)^{1/1+\gamma_*}\ .
\end{equation}
In turn, this implies that Eq.~(\ref{treelevel})
predicts approximate hyperscaling:
\begin{equation}
\label{hypers}
M_\pi\sim M_\tau\sim F_\pi\sim F_\tau\sim M_{\rm h}\sim m^{1/1+\gamma_*}\ ,
\end{equation}
which can also be shown to apply to other hadron masses
generically denoted as $M_{\rm h}$.  dChPT thus predicts that,
in the large-mass regime, the theory behaves approximately as a mass-deformed
conformal theory.  Given the spectral results shown in Fig.~\ref{fig1}, this is the first success of dChPT.   The intuitive
reason for hyperscaling is that, with $m\gg c_1\cal{M}$,
the fermion mass is the dominant source of the breaking of scale invariance.

While all particle masses exhibit hyperscaling,
the pion and the dilaton are still parametrically the lightest
particles in the theory: $M_\pi\sim M_\tau\sim c_1$, which, by
assumption, is a small parameter.   In fact,
\begin{equation}
\label{smallp}
\frac{M_\pi^2}{(4\pi F_\pi)^2}\sim c_1 v(m)\sim c_1\log\frac{m}{c_1\cal{M}}\ ,
\end{equation}
which implies that the expansion underlying dChPT is systematic as
long as $c_1\log\frac{m}{c_1\cal{M}}\ll 1$,
even though $m/{\cal M}$ may be large.
By contrast, ordinary ChPT for QCD is only valid when $m$ is small
relative to the infrared scale of the (massless) theory.

\section{Fits to LSD data \cite{LSD0,LSD} }

\begin{figure}[t]
\begin{center}
\includegraphics*[width=6cm]{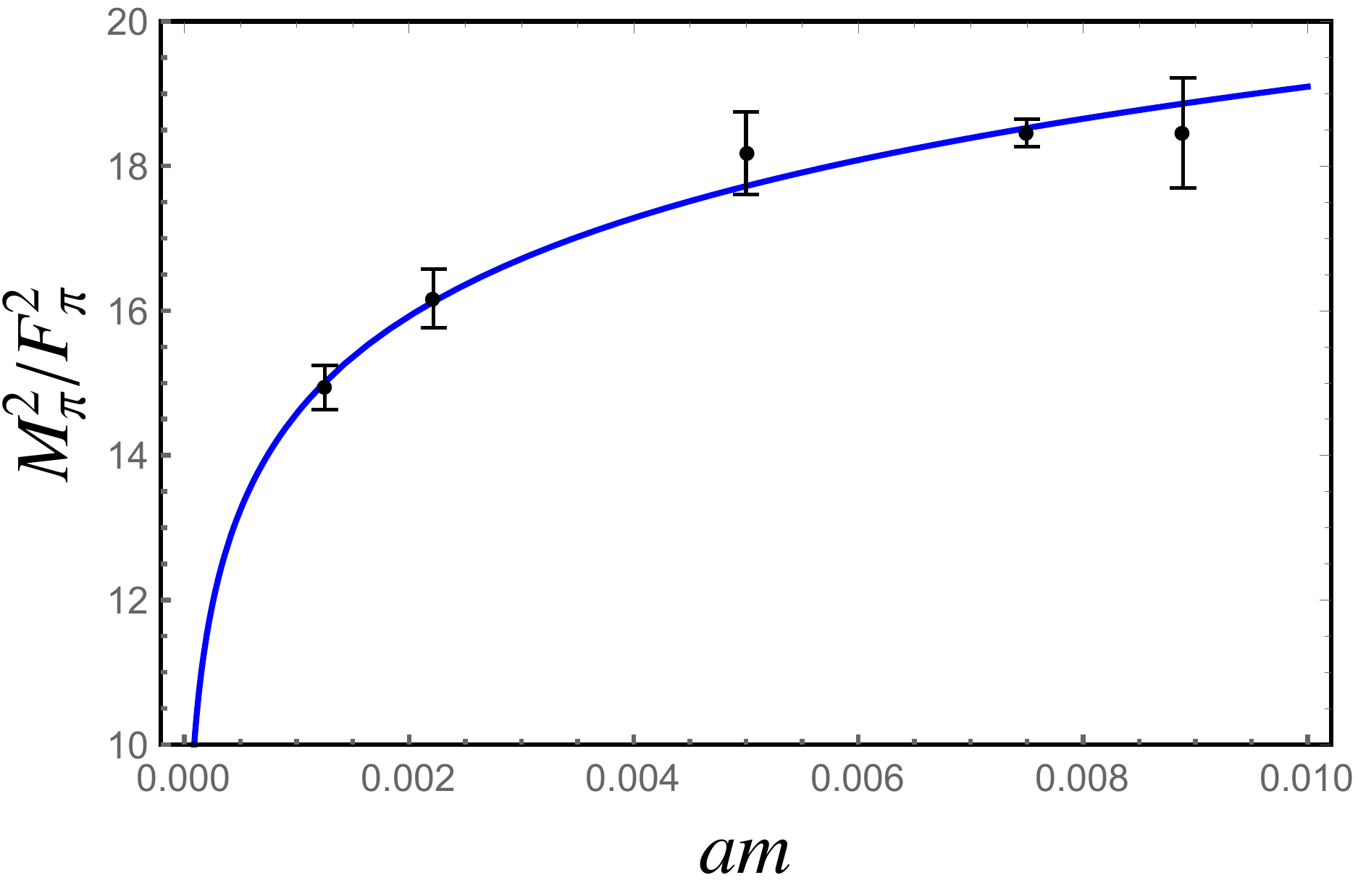}
\hspace{0.5cm}
\includegraphics*[width=6cm]{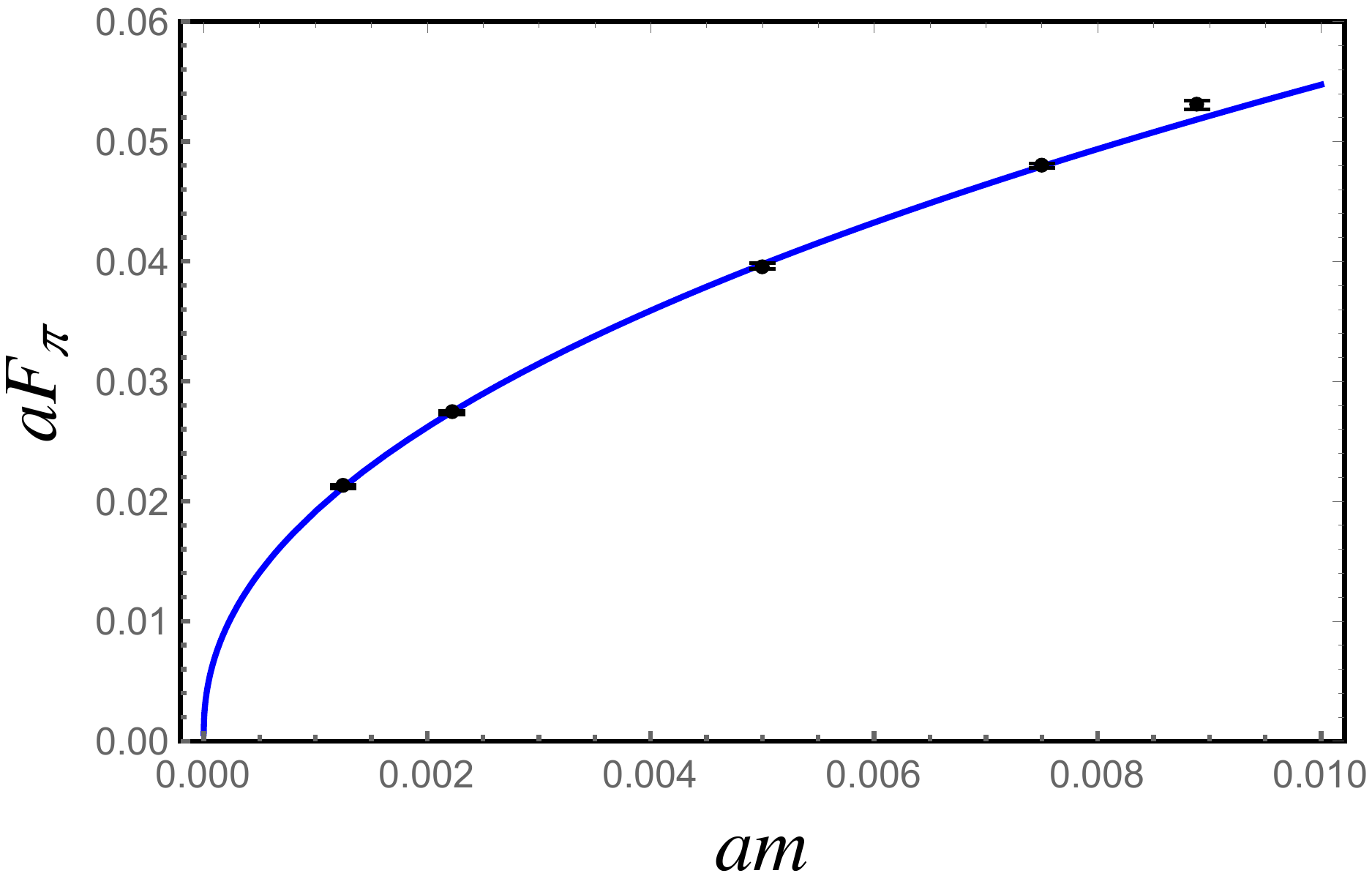}
\hspace{0.5cm}
\includegraphics*[width=6cm]{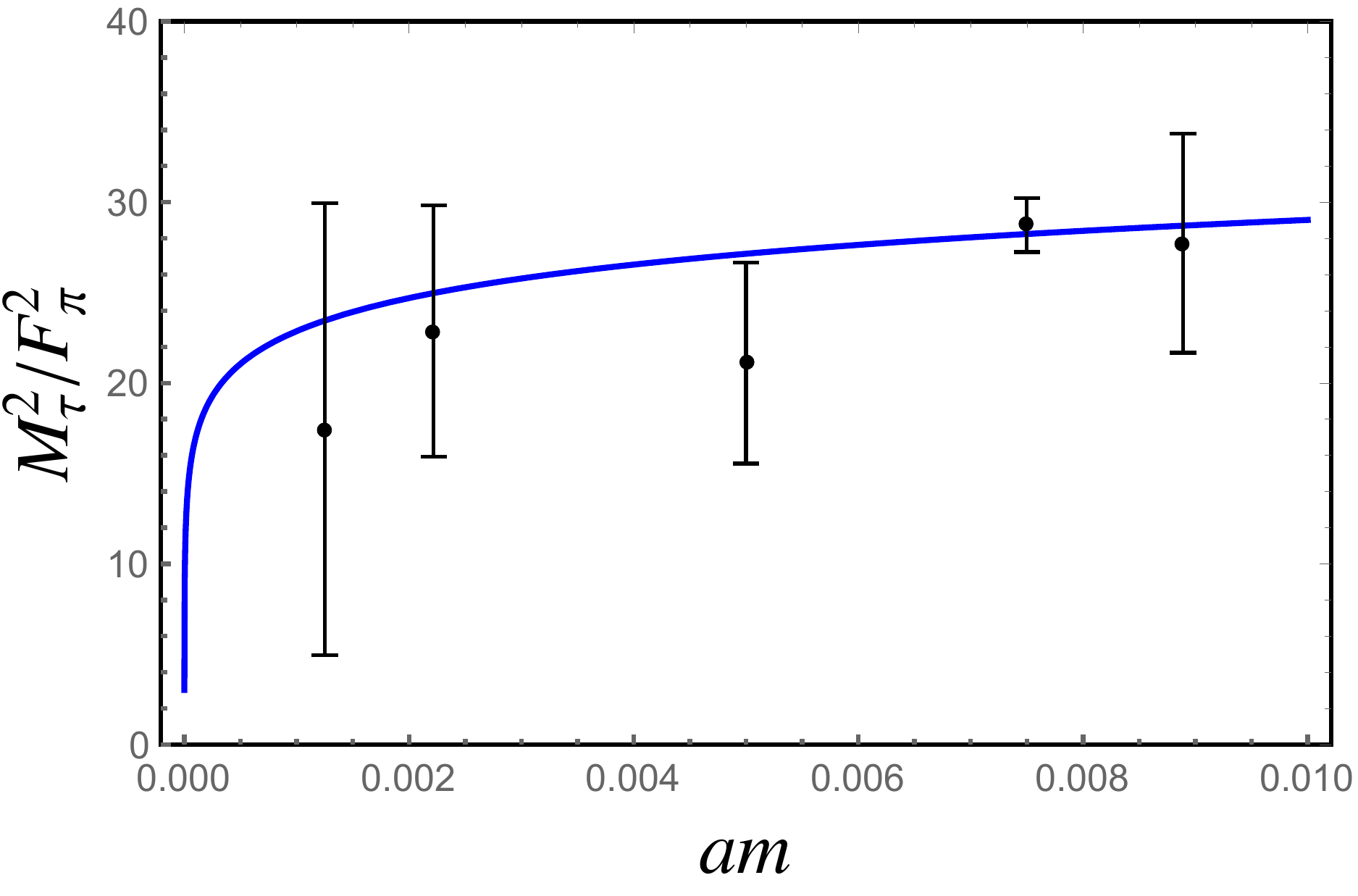}
\caption{\it $M_\pi^2/F_\pi^2$, $aF_\pi$ and $M_\tau^2/F_\pi^2$ as a function of the bare fermion mass in lattice units, $am$.   Taken from Ref.~\cite{GNS}.}
\label{fig2}
\end{center}
\end{figure}

One would like to test dChPT beyond the semi-quantitative
observation of hyperscaling.   Here we review some of the fits performed
in Ref.~\cite{GNS} of the tree-level predictions
to data for the pion mass and decay constant, as well as the dilaton mass,
as a function of the fermion mass.\footnote{
  The dilaton decay constant was not measured in Refs.~\cite{LSD0,LSD,KMI}.}
The ensembles of Ref.~\cite{LSD} all have the same bare coupling,
and fermion masses
$10^3am=(1.25,\ 2.22,\ 5.00,\ 7.50,\ 8.89)$.
In Fig.~\ref{fig2} we show results of one of the fits of Ref.~\cite{GNS}, using data for
$M_\pi^2/F_\pi^2$, $aF_\pi$ and $M_\tau^2/F_\pi^2$
at the lowest four $am$ values.  The fit has a $p$-value of 0.89.
Including also the largest mass leads to a
less good fit, but the $p$-value is still 0.29.
We conclude that tree-level dChPT provides an excellent description
of the LSD data.\footnote{%
  For other applications of tree-level dChPT, see Refs.~\cite{LatHC2,LSD10}.
}

The parameters controlling the mass dependence are well determined
by these fits: we find $\gamma_*=0.94(2)$ and $aB_\pi=2.1(1)$.
In contrast,
finding $af_\pi$ requires a long extrapolation to the chiral limit,
and the result of the fits is $af_\pi=0.0006(3)$.

We emphasize that the data all have $0.02\lesssim aF_\pi(m)\lesssim 0.06$,
with $F_\pi L\gtrsim 1$ and certainly $M_\pi L\gg 1$, so that the data are all in the $p$-regime.   However, as already follows from the
approximate hyperscaling, the LSD fermion masses are solidly in the
large-mass regime, hence the long extrapolation needed
to reach the chiral limit.   This long extrapolation may make the value of $af_\pi$
very sensitive to higher orders in dChPT, even if next-to-leading order (NLO) corrections
at the LSD masses are relatively small.
The (tentative) conclusion is that
reaching the chiral limit while keeping
$f_\pi L\gtrsim 1$ would require unrealistically large volumes,
if the same bare coupling is kept.

The LSD collaboration also measured two of the staggered pion taste splittings
(specifically, the ``axial'' and ``tensor'' ones).
The taste splittings look very different from those found in QCD,
which are explained well by the staggered extension of ordinary ChPT.
Once we extend dChPT to include the
discretization effects of staggered fermions (SdChPT for short),
we find that these taste splittings
can be quantitatively understood using SdChPT.   For details, see
Ref.~\cite{GNS}; we will show an example of taste splittings
in the next section.

\section{Fits to LatKMI data \cite{KMI}}

Next, we consider fitting
dChPT to the data of Ref.~\cite{KMI}.
The LatKMI collaboration considered the same theory, with fermion masses
$10^2am=(1.2,\ 1.5,\ 2.0,\ 3.0,\ 4.0,\ 5.0,\ 6.0,\ 7.0,\ 8.0,\ 10.0)$,
significantly larger than those of the LSD collaboration.
Since the LatKMI and LSD collaborations used different lattice actions,
and both collaborations did simulations at a single value of the bare
coupling, we cannot
compare the two data sets in physical units.
In other words, we do not know the relative size of the
LatKMI and LSD lattice spacings.

In Ref.~\cite{GSKMI} we found that we cannot fit the LatKMI data over
their full mass range with tree-level dChPT.   Fits of subsets suggest
that NLO corrections will have to be taken into account.  However, at NLO
the effective theory has a large number of new LECs \cite{GS},
and there are insufficient data to resolve
these, even if the actual fits contain a smaller
number of linear combinations of the new LECs.

\begin{figure}[t]
\begin{center}
\includegraphics*[width=8cm]{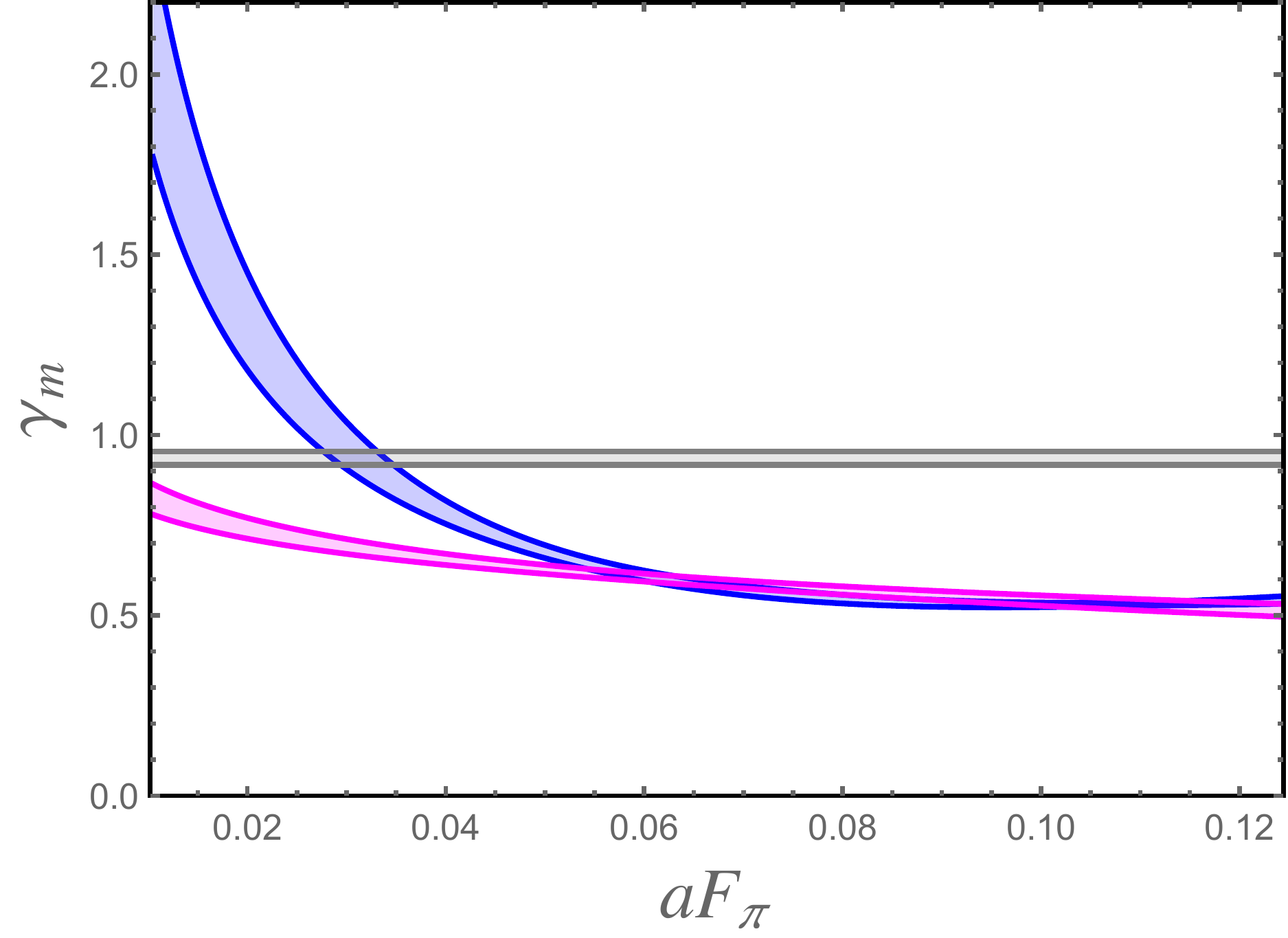}
\caption{\it The running mass anomalous dimension $\gamma(m)$,
obtained from various fits of the LatKMI data,
as a function of $aF_\pi(m)$, see text.
Taken from Ref.~\cite{GSKMI}.}
\label{fig3}
\end{center}
\end{figure}

Instead, we found that the data are well described with
an $m$-dependent mass anomalous dimension\footnote{%
  In Ref.~\cite{GSKMI} we proved that this choice is still consistent with
  the Ward--Takahashi identities for scale invariance.
  One can think of Eq.~(\ref{andim}) as arising from
  a partial resummation of higher orders.
}
\begin{equation}
\label{andim}
\gamma(m)=\gamma_0-bv(m)+cv^2(m)\ .
\end{equation}
For a detailed description of the fits we refer to Ref.~\cite{GSKMI}.
Here we just show the results
we obtain for $\gamma(m)$ of
Eq.~(\ref{andim}), in Fig.~\ref{fig3}.   The blue band shows a fit
($p$-value = 0.48)
to data at all but
the largest mass of the LatKMI range, with
all three parameters, $\gamma_0$,
$b$ and $c$.   The magenta band
shows a fit ($p$-value = 0.19) with the two largest masses omitted, and $c=0$.
For comparison, the gray band
represents $\gamma_*=0.94(2)$ obtained from the LSD data.
The actual data all lie in the range
$0.045\lesssim aF_\pi\lesssim 0.12$, where the two bands overlap, as expected.
Again, the LatKMI masses are in the large-mass
regime, requiring a long extrapolation to the chiral limit.   These
results corroborate the sensitivity of the chiral limit
to higher orders,
even if those higher orders can be small in the range of the data.
Indeed,
small corrections to $v(m)$ over a limited $m$-interval make a larger effect on the ``hyperscaling factor'' $e^{v(m)}$ in the tree-level
expressions~(\ref{treelevel}), which magnifies if $m$ is taken
outside the range of the actual data.

\begin{figure}[t]
\begin{center}
\includegraphics*[width=8cm]{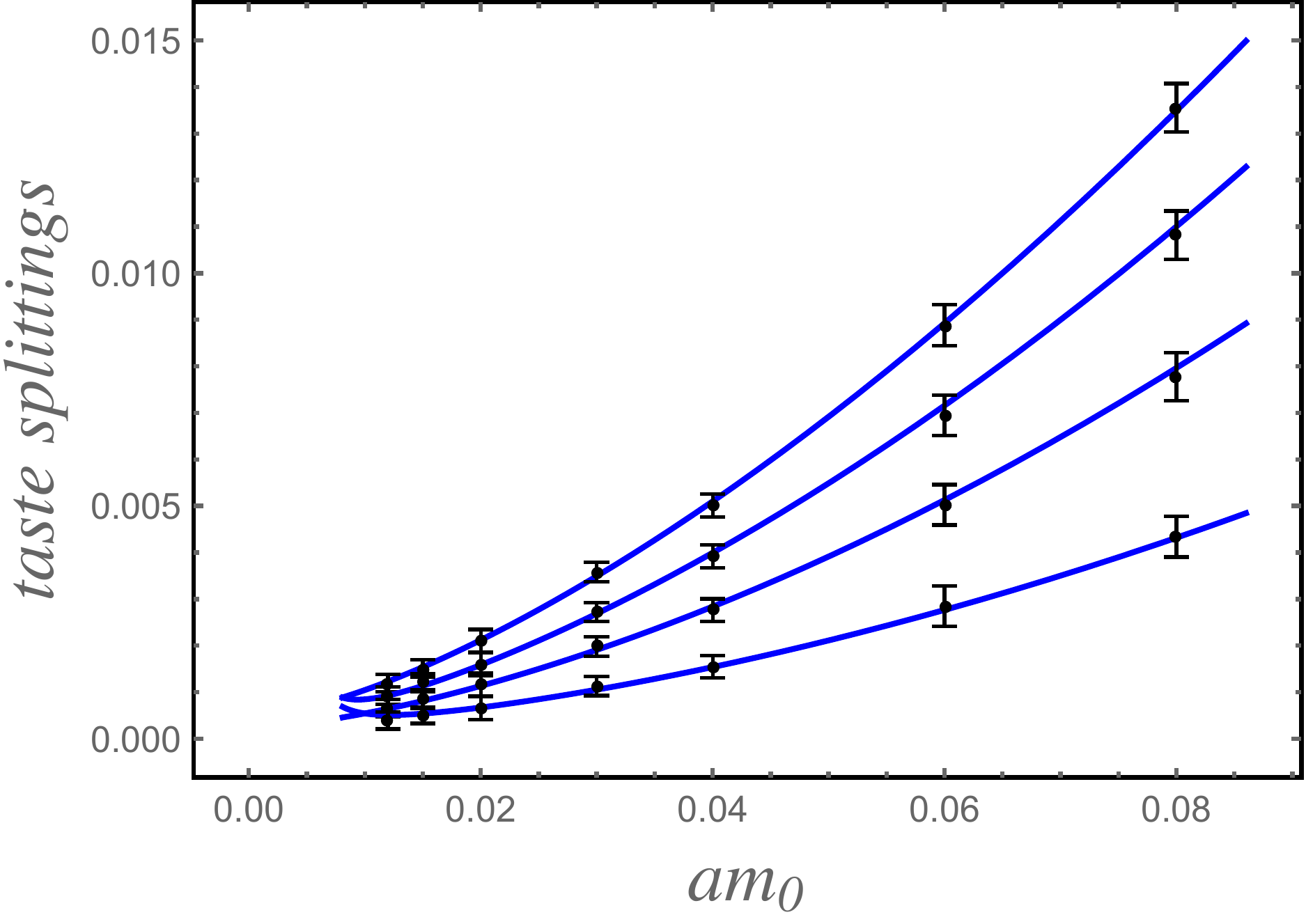}
\caption{\it Fit of tree-level SdChPT to the taste splittings of Ref.~\cite{KMI}.  Taken
from Ref.~\cite{GSKMI}.}
\label{fig4}
\end{center}
\end{figure}

To conclude this section,
we show in Fig.~\ref{fig4} a fit of the taste splittings to
tree-level SdChPT augmented by the varying $\gamma(m)$
of Eq.~(\ref{andim}).%
\footnote{Ref.~\cite{KMI} measured taste splittings only for a subset of seven
fermion masses with a maximum value 0.08.   Setting $c=0$ yields a good fit
($p$-value = 0.44) to tree-level SdChPT
with $\gamma(m)$ of Eq.~(\ref{andim}).}
These taste splittings
behave very different from those in QCD with staggered fermions,
where the taste splittings are essentially independent of $am$.
The goodness of the fit is a
successful test of the dChPT framework.

\section{Conclusion}

We identified a ``large-mass regime'' in dChPT that has no
equivalent in ordinary ChPT. This
is the regime with $c_1{\cal{M}} \ll m \ll c_1{\cal{M}}e^{1/c_1}$.
Much like a mass-deformed infrared-conformal theory,
the large-mass regime of a ``walking''
theory exhibits approximate hyperscaling, while the expansion
underlying dChPT remains systematic, thanks to
the smallness of $c_1\propto |n_f-n_f^*|$,
the distance to the conformal sill.   We found that current simulations
of the $N_f=8$, SU(3) theory are deep in the large-mass regime.

Consequently, reaching the chiral limit from the data
of Refs.~\cite{LSD,KMI} requires a very long extrapolation.
This extrapolation appears to be
very sensitive to the mass range used in the fits, as well as to higher orders in dChPT.   For example, because of this sensitivity,
the chiral-limit value $af_\pi=0.0006(3)$ found from the LSD data
may be afflicted with a larger systematic error than reflected by the
statistical fit error qouted above.
It is thus very difficult to determine whether $af_\pi$ is actually different
from zero!   Likewise, it is difficult to obtain a precise value
for  $c_1{\cal{M}}$.
The difficulty of controlling the long extrapolation raises
the intriguing question whether,
effectively, dChPT may also apply {\it inside} the conformal window,
as long as the theory is mass-deformed.

Both Refs.~\cite{LSD} and
\cite{KMI} measured the dilaton mass, finding it to be
nearly degenerate with the pions. But the errors are so large that the
dilaton mass has
little influence on the fit, as can be seen in Fig~\ref{fig2}.
Given the questions raised above
it would be helpful if more precise data become available,
especially at the lower fermion-mass range.
In particular, this might allow for the probing of NLO corrections.
Also, the size of the taste splittings indicates that scaling violations
are large, and it would be nice to add another lattice spacing to study the
approach to the continuum limit, about which no information
is available at present.

We end with a separate comment.   In Refs.~\cite{AIP,AIP3}, a more
general class of dilaton potentials
\begin{equation}
\label{Delta}
V_\Delta(\tau)\propto\frac{e^{4\tau}}{4-\Delta}\left(1-\frac{4}{\Delta}e^{(\Delta-4)\tau}\right)
\end{equation}
was considered.   For $\Delta\to 4$, one recovers the dilaton
potential in Eq.~(\ref{lag}), while for $\Delta=2$ one
obtains the potential for the $\sigma$-model.   However, as we
pointed out in Ref.~\cite{GSKMI}, for values of $\Delta$ not close
to 4, there is no consistent power counting, and thus the
lagrangian with this generalized potential does not constitute an effective field theory, contrary to what is claimed in Ref.~\cite{AIP3}.

\newpage
\vspace{2ex}
\noindent {\bf Acknowledgments}
\vspace{2ex}

MG's work is supported by the U.S. Department of
Energy, Office of Science, Office of High Energy Physics, under Award
Number DE-SC0013682.
YS is supported by the Israel Science Foundation
under grant no.~491/17.

\end{document}